\documentclass[12pt]{article}
\begin{document}
\title{Revisiting Zitterbewegung}
\author{B.G. Sidharth\\
International Institute for Applicable Mathematics \& Information Sciences\\
Hyderabad (India) \& Udine (Italy)\\
B.M. Birla Science Centre, Adarsh Nagar, Hyderabad - 500 063
(India)}
\date{}
\maketitle
\begin{abstract}
The Dirac wave equation for the electron soon lead to the
recognition of the Zitterbewegung. This was studied both by
Schrodinger and Dirac. Later there were further elegant and
sometimes dissenting insights, from different authors. We briefly
review some of these developments. However in more recent times with
dark energy and noncommutative spacetime coming to centre stage, the
earlier studies of Zitterbewegung become questionable.
\end{abstract}
\section{Introduction}
The phenomenon of Zitterbewegung has been encountered from the very
early days of relativistic Quantum Mechanics. The term itself was
coined by Schrodinger. It shows up in the Dirac theory of the
electron \cite{diracpqm}. Early researchers like Dirac himself and
Schrodinger studied it. Both of them realized that this effect, in
which the Dirac electron appears vibrating with the velocity of
light implies the breakdown of the concept of the measurement of the
velocity instantaneously: The electron vibrates rapidly within the
Compton scale. Our physics starts with averages over the Compton
scale, when we return to sub luminal velocities. In other words the
electron has this high frequency oscillatory motion superimposed
over its normal average motion. If we try to measure the position or
velocity of the electron to better than the Compton scale, indeed we
end up with the creation of electron positron pairs. This shows up
in the complex character of the Dirac coordinates of the electron,
or equivalently the non Hermitian character of the corresponding
operators. So these operators have been considered to be
representing a many particle observable. What this means in a one
particle interpretation is that the trajectory of the electron
cannot be defined by a sharp curve, but rather by a fudged out or
fuzzy tube, whose radius is of the order of the
Compton wavelength. Let us see this in greater detail \cite{barut}.\\
The Hamiltonian for the free electron-positron system according to
Dirac is
\begin{equation}
H = c \vec{\alpha} \cdot \vec{p} + mc^2 \beta ,\label{z1}
\end{equation}
where $\vec{\alpha}$ and $\beta$ are matrices which satisfy the
Dirac algebra
$$\{ \alpha_\imath ,
\alpha_j \} = 2 \delta_{\imath j} \, (\imath, j =
1,2,3),$$
\begin{equation}
\{ \alpha_1 , \beta \} = 0, \, \beta^2 = I.\label{z2}
\end{equation}
The momentum vector $\vec{p}$ and the coordinate vector $\vec{x}$
are taken to satisfy
$$[x_\imath , x_j] = 0 = [p_\imath , p_j],$$
\begin{equation}
[x_\imath , p_j] = \imath \hbar \delta_{\imath j} I,\label{z3}
\end{equation}
and to commute with $\vec{\alpha}$ and $\beta$.\\
In the Heisenberg picture, all these relations hold at any time, and
the time derivative of any one of these operators which do not
depend explicitly on time, say $A$, is given by
\begin{equation}
\frac{dA}{dt} = \imath [H, A]/\hbar\label{z4}
\end{equation}
Thus
\begin{equation}
\frac{d\vec{p}}{dt} = \vec{0}, \frac{dH}{dt} = 0,\label{z5}
\end{equation}
while
\begin{equation}
\frac{d\vec{x}}{dt} = c \vec{\alpha}\label{z6}
\end{equation}
and
\begin{equation}
- \imath \hbar =\frac{d\vec{\alpha}}{dt} = [H, \vec{\alpha}]= -
\{H,\vec{\alpha}\} + 2H\vec{\alpha} = -2c\vec{p} +
2H\vec{\alpha}\label{z7}
\end{equation}
This can be written as
\begin{equation}
- \imath \hbar \frac{d\vec{\alpha}}{dt} = 2H\vec{\eta},\label{z8}
\end{equation}
with
\begin{equation}
\vec{\eta} = \vec{\alpha} - cH^{-1} \vec{p},\label{z9}
\end{equation}
We have
\begin{equation}
-\imath \hbar \frac{d\vec{\eta}}{dt} = -\imath \hbar
\frac{d\vec{\alpha}}{dt} = 2H\vec{\eta}\label{z10}
\end{equation}
so that
\begin{equation}
\vec{n}(t) = e^{2\imath Ht/\hbar_{\vec{\eta_0}}}\label{z11}
\end{equation}
Here $\vec{\eta}_0$ is a constant operator:
\begin{equation}
\vec{\eta}_0 = \vec{\eta}(0) = \vec{\alpha}(0) -
cH^{-1}\vec{p}.\label{z12}
\end{equation}
We can easily check that
\begin{equation}
\{ H,\vec{\eta}\} = 0 = \{ H. \vec{\eta}_0\}\label{z13}
\end{equation}
so that we can also write, from Eq. (\ref{z11}),
\begin{equation}
\vec{\eta}(t) = \vec{\eta}_0 e^{-2\imath Ht/\hbar}.\label{z11'}
\end{equation}
Combining Eqs. (\ref{z7}), (\ref{z9}) and (\ref{z11'}), we obtain
\begin{equation}
\frac{d\vec{x}}{dt} = c \vec{\alpha} = c^2 H^{-1}\vec{p} +
c\vec{n}_0 e^{-2\imath Ht/\hbar}\label{z14}
\end{equation}
which can be integrated again to get
\begin{equation}
\vec{x}(t) = \vec{a} + c^2 H^{-1}\vec{p}t + \frac{1}{2}\imath \hbar
c \vec{\eta}_0 H^{-1}e^{-2\imath Ht/\hbar},\label{z15}
\end{equation}
with $\vec{a}$ a constant (operator) of integration
\begin{equation}
\vec{a} = \vec{x}(0)-\frac{1}{2} \imath \hbar c \vec{\alpha}
(0)H^{-1} + \frac{1}{2}\imath \hbar c^2 H^{-2}\vec{p}.\label{z16}
\end{equation}
Now. Eq. (\ref{z15}) can be written as
\begin{equation}
\vec{x}(t) = \vec{x}_A(t) + \vec{\xi}(t)\label{z17}
\end{equation}
with
\begin{equation}
\vec{x}_A(t) = \vec{a} + c^2 H^{-1} \vec{p}t,\label{z18}
\end{equation}
the form we might expect for the "position" operator of a
relativistic classical point mass. The remaining contribution to
$\vec{x}$ is
\begin{equation}
\vec{\xi}(t)=\frac{1}{2}\imath \hbar c\vec{\eta}_0 H^{-1}e^{-2\imath
Ht/\hbar}= \frac{1}{2} \imath \hbar c \vec{\eta} H^{-1},\label{z19}
\end{equation}
which describes a microscopic, high-frequency Zitterbewegung
superimposed on the microscopic motion (\ref{z18}) associated with
$\vec{x}_A$. The Zitterbewegung has a characteristic amplitude
$\hbar /2mc$, half the Compton wavelength of the electron, and a
characteristic angular
frequency $2mc^2/\hbar$.\\
Nevertheless Zitterbewegung has held the attention of a number of
researchers over the decades and we will now revisit these
interpretations and finally make some comments.
\section{Huang's Approach}
In 1952 K. Huang made an elegant and precise study of Zitterbewegung
\cite{huang}. As this work has been all but forgotten, we reproduce
some important parts of it. We begin by observing that the general
Dirac wave packet may be written in a momentum space expansion of
the form
$$\psi (x,t) = h^{-1} {\bf \int} [C^+ (p) e^{-\imath w t}$$
\begin{equation}
\quad \quad +C^{-} (p)e^{\imath w t}] exp (\imath p \cdot
r/\hbar)d^3 p,\label{e1}
\end{equation}
where $w = \epsilon /\hbar , \epsilon =
+(c^2p^2+m^2c^4)^{\frac{1}{2}}, C^+(p)$ is a linear combination of
"spin up" and "spin down" amplitudes of the free particle Dirac
waves belonging to the momentum {\bf $p$} with positive energy, and
$C^-(p)$ for negative energy. That is,
$$C^+(p) = a_1u_1+a_2u_2$$
\begin{equation}
C^-(p) = a_3u_3 + a_4u_4,\label{e2}
\end{equation}
where the coefficients $a_k$ are complex functions of $p$, and
\begin{equation}
u_1 = \left[\begin{array}{ll} 1\\ 0\\ kp_3\\kp_+
\end{array}\right], \quad u_2 = \left[\begin{array}{ll}
0\\ 1\\ kp_-\\ - l[_3
\end{array} \right],\label{e3}
\end{equation}
$$u_3 = \left[\begin{array}{ll}
-kp_3\\ - kp_+\\ 1 \\ 0
\end{array}\right], \quad u_4 = \left[\begin{array}{ll}
-kp_-\\ kp_3\\ 0\\ 1
\end{array} \right],$$
with
\begin{equation}
k = \frac{c}{\epsilon + mc^2}, \, p \pm = p_1 \pm \imath
p_2.\label{e4}
\end{equation}
The wave packet, (\ref{e2}), is normalized provided we require that
$${\bf \int} [C^{+*}(p)C^+(p) +C^{-*}(p)C^{-}(p)] d^2p = 1.$$
Hence,
$$<t> = c <\alpha> = c {\bf \int} (C^{+*}\alpha C^+ + C^{-*} \alpha
C^-) d^3 p$$
$$ \quad c {\bf \int} [(C^{+*} \alpha C^-) e^{2\imath w t} + (C^{-*}
\alpha C^+) e^{-2\imath wt}] d^2p.$$ the arguments of $C^\pm$ being
understood. Since $\alpha$ is a Hermitian operator, we have
$$<t> = c {\bf \int} (C^{+*} \alpha C^+ + C^{-*} \alpha C^-) d^2 p$$
$$\quad \quad \quad +2c {\bf \int} Re\{ (C^{+*} \alpha
C^-)e^{2\imath wt}\} d^2 p.$$
\begin{equation}
<t> = \bar{v} + 2c {\bf \int} K(p) sin (2wt + \phi (p)) d^2
p,\label{e5}
\end{equation}
where
$$
\bar{v} = c {\bf \int} \frac{p}{mc} \left[ 1 +
\left(\frac{p}{mc}\right)^2\right]^{-\frac{1}{2}} (C^{+*} C^+ -
C^{-*} C^-) d^2 p,$$
$$\quad \quad K(p) = |C^{-*} \alpha C^+ |,$$
\begin{equation}
\phi (p) = tan^{-1} \{ Re(C^{-*}\alpha C^+)/Im(C^{-*} \alpha
C^+)\}.\label{e6}
\end{equation}
Hence
$$<r> = r^0 + \bar{v} t - \lambda {\bf \int} K(p)
[1+(p/mc)^2]^{-\frac{1}{2}}$$
\begin{equation}
\quad \quad \quad \times cos (2 wt + \phi (p)) d^2 p,\label{e7}
\end{equation}
where $\lambda = \hbar /mc$ is the Compton wavelength (divided by
$2\pi$) of the electron.\\
The expectation value for the position is seen to be composed of two
parts; a time-independent part, which describes the average motion
of the electron, and a time-dependent part, which involves
interference terms of the type $C^{+*}\alpha C^-$. It arises from
the interference between the positive and negative energy states of
the electron, and represents the Zitterbewegung of the electron. This is as in
(\ref{z16}).\\
For spin along the $z$ axis, Huang finally deduces:
$$\langle x_1\rangle_\psi = - \lambda sin(2wt + \psi)$$
\begin{equation}
\langle x_2\rangle_\psi = - \lambda cos(2wt + \psi).\label{22}
\end{equation}
We can deduce that while this represents a circulatory motion, the
centre itself is at rest, in this case. Further, the $x$ and $y$
components of $\vec{r} \times \vec{r}^\circ$ vanish on the average
but the $z$ component is proportional to the magnetic moment of the
electron. To sum up:\\
The detailed motion of a free Dirac electron was investigated by
examining the expectation values of the position $\vec{r}$ and of
$\vec{r} \times  \dot{\vec{r}}$ in a wave packet. It was shown by
Huang that the well-known Zitterbewegung may be looked upon as a
circular motion about the direction of the electron spin, with a
radius equal to the Compton wavelength (divided by $2 \pi$) of the
electron. Further the intrinsic spin of the electron may be looked
upon as the "orbital angular momentum" of this motion. The current
produced by the Zitterbewegung is seen to give rise to the intrinsic
magnetic moment of the electron.
\section{The Approaches of Barut and Hestenes}
Schrodinger's proposed "microscopic momentum" vector of the
Zitterbewegung was rejected by Barut in favor of a "relative
momentum" vector, with the value $\vec{P} = c\vec{d}$ in the rest
frame of the center of mass. His oscillatory "microscopic
coordinate" vector was retained in the rest frame, taking the form
$\vec{Q} = \imath (\hbar/2mc)\beta \vec{d}$, and the Zitterbewegung
was described in this frame in terms of $\vec{P}, \vec{Q}$, and the
Hamiltonian $mc^2 \beta$, as a finite three-dimensional harmonic
oscillator with a compact phase space. The Lie algebra generated by
$\vec{Q}$ and $\vec{P}$ is that of $SO(5)$, and in particular
$[Q_\imath , P_j] = - \imath \hbar \delta_{\imath j} \beta$. Barut
argued that the simplest possible finite, three-dimensional,
isotropic, quantum-mechanical system requires such as $SO(5)$
structure, incorporates a fundamental length, and has
harmonic-oscillator dynamics. Dirac's equation was derived as the
wave equation appropriate to the description of such a finite
quantum system in an arbitrary moving frame of reference, using a
dynamical group $SO(3,2)$ which can be extended to $SO(4,2)$. Spin
appears here as the orbital angular momentum associated with the
internal system, and rest-mass energy appears as the internal energy
in the rest frame.\\
Hestenes \cite{hestenes} on the other hand struck a different note.
He observed,\\
"The Zitterbewegung is a local circulatory motion of the electron
presumed to be the basis of the electron spin and magnetic moment. A
reformulation of the Dirac theory shows that the Zitterbewegung need
not be attributed to interference between positive and negative
energy states as originally proposed by Schrodinger. Rather, it
provides a physical interpretation for the complex phase factor in
the Dirac wave function generally. Moreover, it extends to a
coherent physical interpretation of the entire Dirac theory, and it
implies a Zitterbewegung interpretation for the Schrodinger theory
as well."
\section{The Author's Interpretation}
As can be seen from (\ref{z14}), the motion of the Dirac electron
consists of two parts. The first part represents the usual Newtonian
motion of a particle. However, the second part is the very high
frequency Zitterbewegung motion which is superimposed on the
Newtonian motion. The frequency here is of the order of the Compton
frequency that is $\hbar /mc^2$. This motion was elegantly studied
by Huang as described in Section 2. Indeed, as can also be seen from
(\ref{22}), this Zitterbewegung part represents a circulatory motion
with velocity $c$. Indeed the radius of this circuit is of the order
of the Compton wavelength $\lambda$. We can see that the circulation
is given by
\begin{equation}
I = \oint \vec{p} \cdot d\vec{s}\label{29}
\end{equation}
where the circuit has the radius equalling the Compton wavelength
$\lambda = h/2 mc$. Using the fact obtained from (\ref{22}) that $p$
is given by $mc$, the circulation in (\ref{29}) now becomes
\begin{equation}
I = \frac{h}{2}\label{30}
\end{equation}
Equation (\ref{30}) shows that the circulation gives the usual spin
of the electron. The picture that now emerges is that the spinning
electron mimics a vortex which as usual exhibits two motions. The
first is the circulatory motion which represents the electron spin,
and the second is the motion of the vortex as a whole, given by the
Newtonian velocity in (\ref{z14}) \cite{cu,uof,tduniv}.\\
Let us now look at all this in a more recent perspective. we note
that from the realm of Quantum Mechanics the position coordinate for
a \index{Dirac}Dirac particle in conventional theory is given by
\begin{equation}
x = (c^2p_1H^{-1}t) + \frac{\imath}{2} c\hbar (\alpha_1 -
cp_1H^{-1})H^{-1}\label{2De2}
\end{equation}
an expression that is very similar to (\ref{z15}). Infact as was
argued in detail \cite{cu} the imaginary parts of both (\ref{z15})
and (\ref{2De2}) are the same, being
of the order of the \index{Compton wavelength}Compton wavelength.\\
It is at this stage that a proper physical interpretation begins to
emerge. \index{Dirac}Dirac himself observed as noted, that to
interpret (\ref{2De2}) meaningfully, it must be remembered that
Quantum Mechanical measurements are really averaged over the
\index{Compton scale}Compton scale: Within the scale there are the
unphysical \index{Zitterbewegung}Zitterbewegung effects: for a point
\index{electron}electron the velocity equals that of light.\\
Once such a minimum \index{spacetime}spacetime scale is invoked,
then we have a non commutative geometry as shown by
\index{Snyder}Snyder more than fifty years ago \cite{sny1,sny2}:
$$[x,y] = (\imath a^2/\hbar )L_z, [t,x] = (\imath a^2/\hbar c)M_x, etc.$$
\begin{equation}
[x,p_x] = \imath \hbar [1 + (a/\hbar )^2 p^2_x];\label{2De3}
\end{equation}
Moreover relations (\ref{2De3}) are compatible with \index{Special
Relativity}Special Relativity. Indeed such minimum
\index{spacetime}spacetime models were studied for several decades,
precisely to overcome the \index{divergences}divergences encountered
in \index{Quantum Field Theory}Quantum Field Theory
\cite{cu},\cite{sny2}-\cite{fink},
\cite{wolf,leepl}.\\
Before proceeding further, it may be remarked that when the square
of a, which we will take to be the \index{Compton wavelength}Compton
wavelength (including the \index{Planck scale}Planck scale, which is
a special case of the \index{Compton scale}Compton scale for a
\index{Planck}Planck \index{mass}mass viz., $10^{-5}gm$), can be
neglected, then we return to point
\index{Quantum Theory}Quantum Theory and the usual commutative geometry.\\
It is interesting that starting from the \index{Dirac}Dirac
coordinate in (\ref{2De2}), we can deduce the non commutative
geometry (\ref{2De3}), independently. For this we note that the
$\alpha$'s in (\ref{2De2}) are given by
$$\vec \alpha = \left[\begin{array}{ll}
\vec \sigma \quad 0\\
0 \quad \vec \sigma
\end{array}
\right]\quad \quad ,$$ the $\sigma$'s being the \index{Pauli}Pauli
matrices. We next observe that the first term on the right hand side
is the usual Hermitian position. For the second term which contains
$\alpha$, we can easily verify from the commutation relations of the
$\sigma$'s that
\begin{equation}
[x_\imath , x_j] = \beta_{\imath j} \cdot l^2\label{2DeA}
\end{equation}
where $l$ is the \index{Compton scale}Compton scale.\\
There is another way of looking at this. Let us consider the one
dimensional coordinate in (\ref{2De2}) or (\ref{z15}) to be complex.
We now try to generalize this \index{complex coordinate}complex
coordinate to three dimensions. Then we encounter a surprise - we
end up with not three, but four dimensions,
$$(1, \imath) \to (I, \sigma),$$
where $I$ is the unit $2 \times 2$ matrix and $\sigma$s are the
Pauli matrices. We get the special relativistic
\index{Lorentz}Lorentz invariant metric at the same time. (In this
sense, as noted by Sachs \cite{sachsgr}, Hamilton who made this
generalization would have hit upon \index{Special Relativity}Special
Relativity, if he had identified the new fourth coordinate
with time).\\
That is,\\
$$x + \imath y \to Ix_1 + \imath x_2 + jx_3 + kx_4,$$
where $(\imath ,j,k)$ now represent the \index{Pauli}Pauli matrices;
and, further,
$$x^2_1 + x^2_2 + x^2_3 - x^2_4$$
is invariant. Before proceeding further, we remark that special
relativistic time emerges above from the generalization of the
complex one dimensional space coordinate to three dimensions.\\
While the usual \index{Minkowski}Minkowski four vector transforms as
the basis of the four dimensional representation of the
\index{Poincare}Poincare group, the two dimensional representation
of the same group, given by the right hand side in terms of
\index{Pauli}Pauli matrices, obeys the quaternionic algebra of the
second rank
\index{spin}spinors (Cf.Ref.\cite{bgsfpl162003,shirokov,sachsgr} for details).\\
To put it briefly, the \index{quarternion}quarternion number field
obeys the group property and this leads to a number system of
quadruplets as a minimum extension. In fact one representation of
the two dimensional form of the \index{quarternion}quarternion basis
elements is the set of \index{Pauli}Pauli matrices. Thus a
\index{quarternion}quarternion may be expressed in the form
$$Q = -\imath \sigma_\mu x^\mu = \sigma_0x^4 - \imath \sigma_1 x^1 - \imath \sigma_2 x^2 -
\imath \sigma_3 x^3 = (\sigma_0 x^4 + \imath \vec \sigma \cdot \vec
r)$$ This can also be written as
$$Q = -\imath \left(\begin{array}{ll}
\imath x^4 + x^3 \quad x^1-\imath x^2\\
x^1 + \imath x^2 \quad \imath x^4 - x^3
\end{array}\right).$$
As can be seen from the above, there is a one to one correspondence
between a \index{Minkowski}Minkowski four-vector and $Q$. The
invariant is now given by
$Q\bar Q$, where $\bar Q$ is the complex conjugate of $Q$.\\
However, as is well known, there is a lack of
\index{spacetime}spacetime reflection \index{symmetry}symmetry in
this latter formulation. If we require reflection
\index{symmetry}symmetry also, we have to consider the four
dimensional representation,
$$(I, \vec \sigma) \to \left[\left(\begin{array}{ll}
I \quad 0 \\
0 \quad -I
\end{array}\right), \left(\begin{array}{ll}
0 \quad \vec \sigma \\
\vec \sigma \quad 0
\end{array}\right)\right] \equiv  (\Gamma^\mu)$$
(Cf.also.ref. \cite{heine} for a detailed discussion). The
motivation for such a reflection \index{symmetry}symmetry is that
usual laws of physics, like \index{electromagnetism}
electromagnetism do indeed show the symmetry.\\
We at once deduce \index{spin}spin and \index{Special
Relativity}Special Relativity and
 the geometry (\ref{2De3}) in these considerations. This is a transition that has
 been long overlooked \cite{bgsfpl152002}.
 It must also be mentioned that \index{spin}spin half itself is relational and refers to
 three dimensions, to a \index{spin}spin network infact \cite{penrose,mwt}. That is,
 \index{spin}spin half is not meaningful in a single particle \index{Universe}Universe.\\
While a relation like (\ref{2DeA}) above has been in use recently,
in non commutative models, we would like to stress that it has been
overlooked that the origin of this non commutativity
lies in the original \index{Dirac}Dirac coordinates. (Conversely, starting with
the noncommutative geometry (\ref{2DeA}), we can argue that the Dirac equation
can be obtained (Cf.ref.\cite{tduniv})).\\
The above relation shows on comparison with the position-momentum
commutator that the coordinate $\vec x$ also behaves like a
``momentum''. This can be seen directly from the \index{Dirac}Dirac
theory itself where we have \cite{gueret}
\begin{equation}
c\vec \alpha = \frac{c^2\vec p}{H} - \frac{2\imath}{\hbar} \hat x
H\label{2Dea}
\end{equation}
In (\ref{2Dea}), the first term is the usual momentum. The second
term is the extra
``momentum'' $\vec p$ due to \index{Zitterbewegung}Zitterbewegung we have
already encountaered.\\
Infact we can easily verify from (\ref{2Dea}) that
\begin{equation}
\vec p = \frac{H^2}{\hbar c^2}\hat x\label{2Deb}
\end{equation}
where $\hat x$ has been defined in (\ref{2Dea}).\\
We finally investigate what the angular momentum $\sim \vec x \times
\vec p$ gives - that is, the angular momentum at the \index{Compton
scale}Compton scale. We can easily show that
\begin{equation}
(\vec x \times \vec p)_z = \frac{c}{E} (\vec \alpha \times \vec p)_z
= \frac{c}{E} (p_2\alpha_1 - p_1\alpha_2)\label{2Dec}
\end{equation}
where $E$ is the eigen value of the Hamiltonian operator $H$.
Equation (\ref{2Dec}) shows that the usual angular momentum but in
the context of the minimum \index{Compton scale}Compton scale cut
off, leads to the ``mysterious'' Quantum Mechanical \index{spin}spin. This
resembles considerations in the previous section.\\
In the above considerations, we started with the \index{Dirac}Dirac
equation and deduced the underlying non commutative geometry of
spacetime.\\
In recent years, dark energy or the vacuum Zero Point Field has
occupied centre stage. Indeed the author's 1997 work pointed to a
dark energy driven accelerating universe with a small cosmological
constant and this was dramatically confirmed the very next year
(Cf.ref.\cite{cu} and references therein for details).\\
We would first like to point out that a background Zero Point Field
of this kind can explain the Quantum Mechanical spin half (as also
the anomalous $g = 2$ factor) for an otherwise purely classical
electron \cite{sachi,uof,milonniqv}. The key point here is
(Cf.ref.\cite{sachi}) that the classical angular momentum $\vec r
\times m \vec v$ does not satisfy the Quantum Mechanical commutation
rule for the angular momentum $\vec J$. However when we introduce
the background Zero Point Field (ZPF), the momentum now becomes
\begin{equation}
\vec J = \vec r \times m \vec v + (e/2c) \vec r \times (\vec B
\times \vec r) + (e/c) \vec r \times \vec A^0 ,\label{3ez5}
\end{equation}
where $\vec A^0$ is the vector potential associated with the ZPF and
$\vec B$ is an external magnetic field introduced merely for
convenience, and which can be made vanishingly
small.\\
It can be shown that $\vec J$ in (\ref{3ez5}) satisfies the Quantum
Mechanical commutation relation for $\vec J \times \vec J$. At the
same time we can deduce from (\ref{3ez5})
\begin{equation}
\langle J_z \rangle = - \frac{1}{2} \hbar
\omega_0/|\omega_0|\label{3ez6}
\end{equation}
Relation (\ref{3ez6}) gives the correct Quantum Mechanical results referred to above.\\
From (\ref{3ez5}) we can also deduce that
\begin{equation}
l = \langle r^2 \rangle^{\frac{1}{2}} =
\left(\frac{\hbar}{mc}\right)\label{3ez7}
\end{equation}
Equation (\ref{3ez7}) shows that the mean dimension of the region in
which the ZPF fluctuation contributes is of the order of the Compton
wavelength of the electron. By relativistic covariance
(Cf.ref.\cite{uof}), the corresponding time scale is at the Compton
scale. As noted, Dirac, in his relativistic theory of the electron
encountered the Zitterbewegung effects within the Compton scale
\cite{diracpqm} and he had to invoke averages over this scale to
recover meaningful results. We have also shown earlier, without
invoking the ZPF \cite{bgsfpl152002} how spin follows, that is we
get (\ref{3ez6}) and (\ref{3ez7}) using Zitterbewegung.\\
In any case, the earlier precise characterization of Zitterbewegung
assumed that time and space can be precisely characterized at
arbitrarily small scales. This assumption does not stand the
scrutiny of later work \cite{wigner,tduniv}.

\end{document}